\documentclass[aps,prl,twocolumn,preprintnumbers,
superscriptaddress,floatfix]{revtex4}

\pdfoutput=1

\usepackage{amssymb,multirow}
\usepackage{graphicx}

\begin{document}

\newcommand{\Tr}{\mbox{Tr\,}}
\newcommand{\beq}{\begin{equation}}
\newcommand{\eeq}{\end{equation}}
\newcommand{\bea}{\begin{eqnarray}}
\newcommand{\eea}{\end{eqnarray}}
\renewcommand{\Re}{\mbox{Re}\,}
\renewcommand{\Im}{\mbox{Im}\,}

\voffset 1cm

\newcommand\sect[1]{\emph{#1}---}

\pagestyle{empty}

\preprint{
\begin{minipage}[t]{3in}
\begin{flushright} SHEP-09-15
\\[30pt]
\hphantom{.}
\end{flushright}
\end{minipage}
}

\title{Hadronization at the AdS Wall}

\author{Nick Evans$^1$, James French$^1$, Kristan Jensen$^2$ \& Ed Threlfall}

\affiliation{ School of Physics and Astronomy, University of
Southampton, Southampton, SO17 1BJ, UK \\ \vspace{0.2cm}  $^2$
Department of Physics, University of Washington, Seattle, WA98195,
USA \\ \vspace{0.2cm} evans@soton.ac.uk, james.french@soton.ac.uk,
kristanj@u.washington.edu, ejt@phys.soton.ac.uk}

\begin{abstract}
\noindent We describe hadronization events, using the AdS/CFT
Correspondence, which display many of the qualitative features
expected in QCD. In particular we study the motion of strings with
separating end points in a back-reacted hard wall geometry. The
solutions show the development of a linear QCD-like string. The
end points oscillate in the absence of string breaking. We
introduce string breaking by hand and evolve the new state
forward in time to observe the separation of two string segments.
A kink associated with this breaking evolves to the end points of
the string inducing rho meson production. We explicitly compute
the rho meson production at the end point.
\end{abstract}

\maketitle

\section{Introduction}

The AdS/CFT Correspondence
\cite{Malda,Witten:1998qj,Witten:1998b,Gubser:1998bc} provides a
new tool for the study of strongly coupled gauge theories and has
been extended to cover both gauge and quark
\cite{Polchinski,Bertolini:2001qa,Kirsch,Karch,
Mateos,Babington,Ghoroku:2004sp,Erdmenger:2007cm} degrees of
freedom. Hadronization, the conversion of quark pair production
into jets of hadrons, is traditionally a thorny problem in QCD. It
is therefore interesting to model this process with the
Correspondence.

The combination of asymptotic freedom and confinement in QCD
ensures that the road from quark production to jets, as in a
high-energy accelerator such as LEP, is broken up into three
overlapping stages. This is the statement of factorization. For a
recent review, see \cite{Ellis:2007ib}. The first, pair
production, typically occurs at high energies and so the relevant
degrees of freedom are a bare quark/anti-quark pair. As they fly
apart, these quarks begin to radiate soft gluons and
quark/anti-quark pairs - the process known as showering - so that
the relevant degrees of freedom are \emph{dressed} quarks.
Showering can be computed at high energies in QCD with effective
theories including soft collinear effective theory. After the
dressed quark/anti-quark pair have propagated a distance of order
$1/\Lambda_{\rm QCD}$ apart, the strong coupling and confining
dynamics of QCD become relevant: the quark pair forms a highly
excited colourless hadron that can and does fragment. This last
conversion is known as hadronization.

The middle process of showering has been studied in different
contexts via the Correspondence. A number of authors
\cite{Alday:2007he,Hatta:2008tx} have computed the evolution after
the injection of $R$ charge into the ${\cal N}=4$ Super Yang-Mills
theory. The ${\cal N}=4$ theory is strongly coupled at all scales
and so there is no suppression of the emission of large transverse
momentum - rather than seeing jets, the events fill the entire
space surrounding the initial pair creation.

In the AdS/CFT correspondence, strings with their ends tied to a
D7 brane represent a quark anti-quark pair in the dual field
theory.
The dual quarks have a constituent mass that  goes to zero in the
limit that the D7 brane fills the entire AdS space. The evolution
of such pairs in the massless quark limit with separating end
points was first studied in \cite{Chesler:2008wd}. They worked in
the quenched ${\cal N}=2$ gauge theory obtained by placing probe
D7 branes in AdS$_5 \times S^5$. Both the endpoints and the string
between them fall indefinitely into the radial direction of the
AdS space representing the spreading of baryon charge and energy
respectively in the dual field theory as the state evolves. These
states are anisotropic and their evolution naturally describes the
showering of flavour.
Hadron (rho meson) production can only occur on the D7 world-volume
through the motion of the endpoints of the string. The production for these states
was computed \cite{Chesler:2008wd} and jet-like structures were
seen.

Much attention has focused on
heavy quark propagation at finite temperature
\cite{Herzog:2006gh,Gubser:2006bz}. The ${\cal N}=4$ theory is
perhaps more like QCD in its non-conformal high temperature phase
which is described by an AdS-Schwarzschild geometry
\cite{Witten:1998b}. These computations have also been inspired by
attempts to reproduce jet quenching observed in heavy ion
collision data at RHIC.

In this paper we wish to return to study the second and third
phases of hadronization described above at zero temperature. To
find a dual gravity description that behaves like QCD one must
seek a non-conformal version of the AdS/CFT Correspondence that at
least incorporates confinement. There has been a considerable body
of work in this direction. We will work in the context of deformed
AdS geometries
\cite{Girardello:1998pd,Pilch:2000ue,Girardello:1999bd,Pilch:2000fu,
Polchinski:2000uf,Babington:2002qt,Kehagias:1999tr,Gubser:1999pk,Constable:1999ch}
which have the benefit of being 3+1 dimensional at all energy
scales and having naive perturbative dimensions for a class of
operators
 in the UV (of course the UV is not perturbative but
conformal, strongly coupled, and highly supersymmetric). Simple
deformations such as the introduction of masses for the ${\cal
N}=4$ matter content that break the supersymmetry to
$\mathcal{N}=2$ \cite{Pilch:2000ue}, $\mathcal{N}=1$,
\cite{Girardello:1999bd,Pilch:2000fu, Polchinski:2000uf} or less
\cite{Babington:2002qt,Kehagias:1999tr,Gubser:1999pk,Constable:1999ch} have been
studied. These dualities are not as well understood as the
original Correspondence but a number of features seem generic. In
particular, the geometries dual to confining theories develop a
wall structure in the interior that stops fields from penetrating
to energy scales (radial distances in AdS) below the mass gap of
the theory. Such a repulsive wall leads to linear potentials
between quarks since they are described by the ends of a
fundamental string in the geometry - typically the centre of these
strings fall on to and lie close to the repulsive wall with their
energy then just scaling with their length.

In fact, the imposition of a simple hard wall cut off on AdS is now
frequently used to model a mass gap
\cite{Polchinski:2001tt,Erlich:2005qh,Da Rold:2005zs}. That will
not be sufficient for us since we wish to evolve the strings onto
the wall and therefore need the repulsion to be described by the geometry.
Given that these dualities are not fully understood we
will take the simplest example of a deformation with such a hard
wall. So called ``dilaton flow" geometries
\cite{Kehagias:1999tr,Gubser:1999pk,Constable:1999ch} have a non-trivial dilaton
profile in a deformed AdS space. The dilaton, or on the field
theory side the gauge coupling, blows up at some infra-red scale
generating the wall (effectively the scale $\Lambda_{\rm QCD}$).
The wall is repulsive to strings and also to the D7 branes they
are attached to, which dynamically induces both confinement
\cite{Brevik:2005fs,Gubser:1999pk} and chiral symmetry breaking
\cite{Babington,Ghoroku:2004sp}. Such a geometry is clearly a
decent first stab at describing a QCD-like gauge theory.

Having found a model for such a theory, we follow the program of
\cite{Chesler:2008wd} and evolve strings in this geometry. In
particular, we consider numerical solutions of the Polyakov string
action for strings with ends tied to a D7 brane and point-like
initial conditions to represent the pair creation of a quark
anti-quark pair. We allow the end points to separate quickly
leaving a string falling into the geometry between. It is fairly
easy (though computer-time consuming) to numerically follow this
evolution with pre-built numerical integrator functions such as
NDSolve in Mathematica. As one might expect, in the hard wall
setup the string falls onto and bounces off the wall in the radial
direction - the centre of the string then oscillates, falling on
to the wall repeatedly. Whilst this happens, the endpoints move
apart and a longer string grows, taking up the kinetic energy near
the end points. Formally, we study this process at infinite $N$
and so there is no string breaking; instead, the string continues
to evolve.
Since the string can only extend so far, the endpoints eventually
stop and their motion reverses. The string bounces in the three
dimensional space. The model therefore correctly reproduces
expectations from field theory in the large $N$ limit.

For real QCD string breaking is presumably greatly favoured,
occurring essentially as soon as there is of order $\Lambda_{\rm
QCD}$ worth of potential energy in the string. In our gravity
description, string breaking is captured by stringy $1/N$ effects:
there will be some transition amplitude between our initial string
and a final state and the actual evolution will involve a sum over
all possible intermediate states, including those with two string
segments. Attempts have been made to compute this breaking probability
in \cite{break1,break2,break3,break4} which we do not add to here.
We choose instead to study the evolution after string breaking,
so we simply break our strings by hand and consider
particular states with two string segments. In particular, we
insert a static quark/anti-quark pair and then evolve these new
\emph{particular} strings. We simulate the inserted pair with a
static string that falls straight down into the AdS space from the
D7 brane. At a time of our choosing (usually once the initial
string grows to a size of order $1/\Lambda_{\rm QCD}$), we break
the initial string in the middle by hand and join each half to a
static segment. We then follow this broken string as it evolves.


This configuration again evolves as one might expect - the static
endpoint (anti-quark) is dragged by the fast moving endpoint
(quark) that it has been attached to and the endpoints
(quark/anti-quark pair) of the original string are free to
separate to infinity. The broken strings have a kink at the point
where the static string was attached which evolves to the
endpoints. The kink then causes a rapid jerk in the motion of the
inserted static endpoint (anti-quark).

In the AdS/CFT Correspondence, mesons made of the quarks are
associated with open strings tied to the D7 brane \cite{Mateos} -
in fact, in the large $N$ and strong coupling limit, rho mesons are the dominant
such mode, described by a dual gauge field on the D7 world volume. The
string endpoint is a source for that gauge field. The rapid jerk
of an endpoint provides a mechanism for copious production of the
rho mesons, in the same way that an accelerating charge
radiates in classical electromagnetism. We show how to compute this production.

The rho meson and each of its radially excited states is
associated with a normalizable mode in the radial direction on the
D7 brane. The full production of the gauge field on the D7
world-volume can therefore be computed mode by mode. In fact the
radial dynamics simply encodes the mass of the physical state and
the strength with which a particular source couples to that mode.
For a single mode, the calculation is just of the usual form for a
moving source coupled to a massive vector field in 4d and is
straightforwardly computed. A static quark (string endpoint) has
a cloud of mesons around it with their density decaying with
distance as an exponential of the mass. When an endpoint accelerates, it
also radiates waves which correspond to the production of rho
mesons in the hadronization event. We explore this production for the rho meson and its excited states.

The solutions we present for moving string solutions provide a
qualitative understanding of how hadronization may occur in QCD or
related theories. We do not in this paper wish to make any claims
that these computations are numerically accurate for QCD, but
 we hope that the mechanisms revealed will
provide thought for the future modelling of hadronization in QCD.
We discuss the prospects in our final section.

\begin{centering}
\includegraphics[width=80mm]{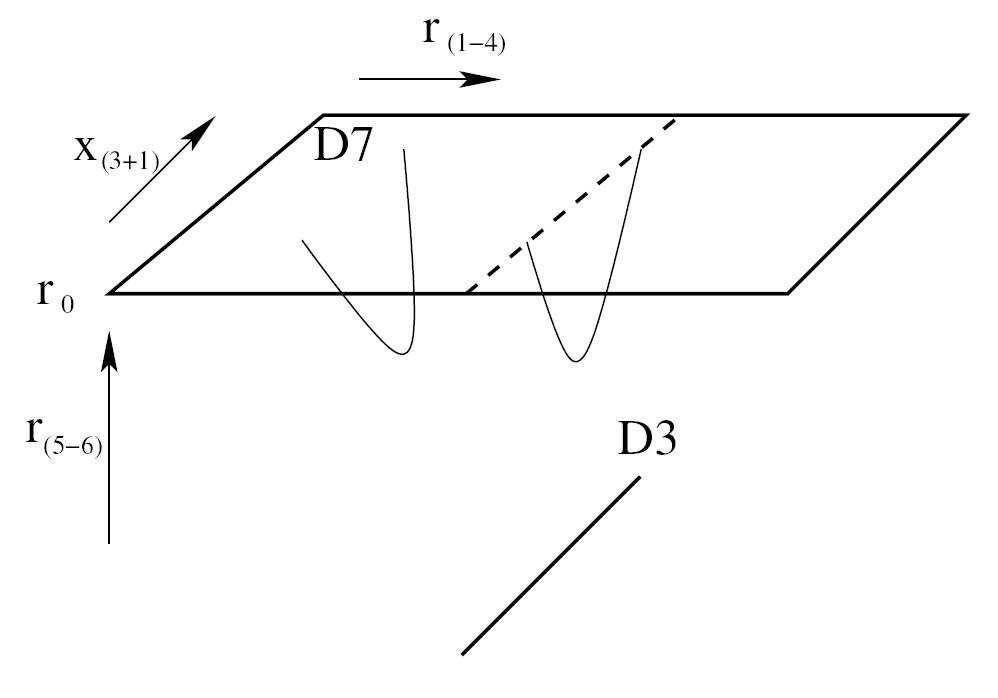}
\end{centering}
{\small{ \label{D3D7fig} Figure 1: The basic D3-D7 system showing
the coordinate labelling we use - note the D3 lies at $r=0$. We
also show two possible moving string configurations - one lies off
the plane of closest approach between the D3 and the D7 and lies
in a 3d space, the string with end points separating along that
plane lies in just 2d. We study moving strings of the latter
type.}}
\par \vspace{-0.4cm}

\section{Geometries and Quarks}

The vacuum structure of ${\cal N}=4$ $SU(N)$ super Yang-Mills gauge theory is
encoded in the AdS$_5\times S^5$ geometry \beq ds^2 = {r^2 \over
R^2} dx_{3+1}^2 + {R^2 \over r^2} dr_6^2 \eeq where $R^4 = 4 \pi
g_sN \alpha'^{2}$, $g_s$ is the string coupling, and $\alpha'$ sets the string length.
The dilaton is constant and there is a four
form $C_{(4)} =  r^4/R^4 dx^0 \wedge..dx^3$. Note that, in these
coordinates, $r$ (the radial direction in the six dimensional $r_6$
space) corresponds to the  gauge theory energy scale ($r=0$ is the infra-red).
The AdS space has a boundary at $r=\infty$ and the dual gauge theory
``lives" on that boundary with a metric conformal to the Minkowski space metric $dx_{3+1}^2$.

To introduce a single quark flavour (really a single
$\mathcal{N}=2$ hypermultiplet), we place a D7 brane lying in the
$x^0-x^3$ and $r^1-r^4$ directions (see Figure 1). In pure AdS,
such a brane lies flat with $r_5^2 + r_6^2= r_0^2$. The separation
$r_0$ sets the scale of the constituent quark mass ($m_q = r_0/2
\pi \alpha'$), related to the energy of an open string with one
end at the bottom of the D7 brane and the other at the AdS horizon
$r=0$. In contrast, a fundamental string with each end on the D7
brane represents a quark/anti-quark pair and the interaction
energy between them. As shown in \cite{Maldacena:1998im}, for very
heavy static quarks (i.e. with their ends tied to a D7 brane at
$r=\infty$), the dual string hangs further and further into the
bulk as the quarks are separated. The total energy of the state
scales as one over the quark separation.

To provide a gravity description of a confining gauge theory, we
will consider a geometry that is asymptotically AdS (the UV
contains the same degrees of freedom as the ${\cal N}=4$ theory)
but which has a back-reacted hard wall in the IR. The simplest
such example we know is a dilaton controlled deformation
\cite{Kehagias:1999tr,Gubser:1999pk}. 
The metric in Einstein frame for this deformed geometry is
\begin{equation} \label{wallmetric}
ds^2 =  {r^2 \over R^2}  A^2(r) dx_{3+1}^2 +  {R^2 \over r^2} dr_6^2,
\end{equation} with
\begin{equation}
A(r) =  \left(1 - ({r_w \over r})^8\right)^{1/4} , ~~~~~ e^\phi =
\left( {1+ (r_w/r)^4  \over 1 - (r_w/r)^4} \right)^{\sqrt{3/2}},
\end{equation} and the four form is as in the pure AdS solution.

Note that the dilaton (dual to the gauge coupling) and metric have
a singularity at $r=r_w$ which can be loosely interpreted as the
scale at which the gauge coupling diverges, $\Lambda_{\rm QCD}$.
In \cite{Evans:2008tu} it was argued that the best interpretation
of this geometry might be as dual to the low energy theory below
some UV cut off where some highly irrelevant supersymmetry
breaking but $SO(6)$ preserving coupling becomes important in the
${\cal N}=4$ gauge theory (a scalar mass term is an example of
such an operator).  For our purposes in this paper though, the
geometry is simply playing the role of a back reacted hard wall to
include confinement and the precise physics is unimportant.

\begin{centering}
\includegraphics[width=80mm]{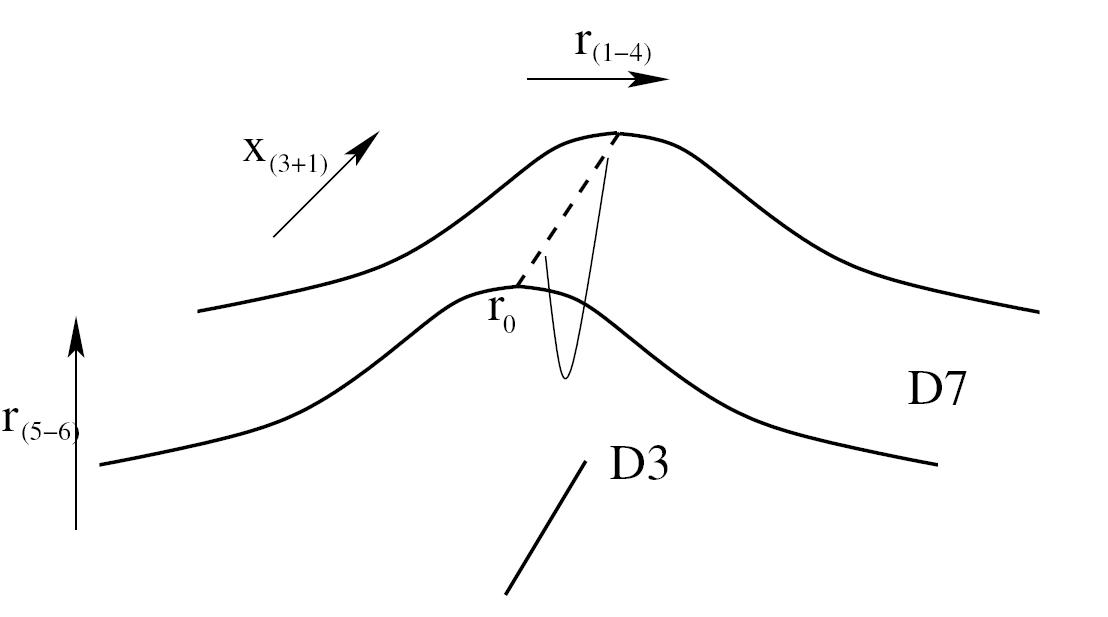}
\end{centering}
{\small{
\label{backD3D7fig}
Figure 2: A sketch of a D7 embedding in the back reacted
hard wall geometry for a quark with zero current mass. The bending
of the D7 induces a constituent mass (non-zero minimum D3-D7
separation) for the quarks. }}
\par

\newpage

Quarks can generically be added to the theory by including a D7
brane as in the AdS case. The hard wall is repulsive to the D7
brane and so induces chiral symmetry breaking through the
formation of a quark condensate - the details are discussed in
\cite{Babington,Ghoroku:2004sp}. In Figure 2 we sketch the form of
the D7 embedding in the geometry for the case of zero bare quark
mass - the D7 is bent so that there is never a zero length string
between the D3 and D7: the quarks have acquired a constituent
mass.

As discussed in \cite{Brevik:2005fs,Gubser:1999pk} a string with
ends tied to a D7 brane, corresponding to a quark pair, is
repelled by the hard wall at $r=r_w$ and the string lies close to
that wall - there is a linear potential between the quarks.

\subsection{Conventions}
Throughout this work, we consider fields living on four different
spaces: the full 10d geometry, the worldvolume of a D7 brane, the
worldsheet of a fundamental string, and on the Minkowski spacetime
of the dual field theory. To simplify some of this chaos, we
choose some conventions for how we label indices and these fields.
First, we will use the upper-case latin letters ($M,N,..$) to
refer to 10d spacetime indices (e.g. $G_{MN}dx^{M}dx^{N}$ is the
metric on the 10d space), while lower-case latin letters
($a,b,..$) refer to worldsheet and worldvolume indices. Finally,
we use lower-case Greek letters ($\mu,\nu,..$) to index the
Minkwoski directions in both the 10d geometry and D7 worldvolume
(i.e. $x^{\mu}$, $\mu=0,1,2,3$).

\section{Strings in Pure AdS}

A study of string evolution in pure AdS was presented in
\cite{Chesler:2008wd}. Here we repeat that framework, which we
will use below, and summarize the form of the results.

We work with the Polyakov action in Einstein frame \beq S_P = - {1
\over 4\pi \alpha'} \int d^2 \sigma e^{\phi/2} \sqrt{-\eta} \eta^{ab}
\partial_a X^M
\partial_b X^N G_{MN}, \eeq where $\eta_{ab}$ is the world sheet
metric, $\sigma^a$ labels the worldsheet coordinates, $X^{M}$
are the embedding functions, and $G_{MN}$ is the background metric.
One is free to use the symmetries of the theory to fix the
world sheet metric and we will pick the form
\beq \eta_{ab} = \left( \begin{array}{cc} - \Sigma(x,r) & 0 \\
0 & {1 \over \Sigma(x,r)} \end{array}\right). \eeq We pick and
choose the form of the stretching function $\Sigma$ problem by
problem in order to keep our numerics stable. Also, we label  the
timelike world sheet coordinate as $\tau$ and the spacelike as
$\sigma$ (i.e. $d^2\sigma=d\tau d\sigma$).

The equations of motion are \beq \label{eom} \begin{array}{l}
\partial_a [ e^{\phi/2}\sqrt{-\eta} \eta^{ab}
G_{MN}
\partial_b X^N] \\ \\ \left. \right.~~~ - {1 \over 2}  \partial_a X^P
\partial_b X^N{\partial \over \partial
X^M}\left( e^{\phi/2}\sqrt{-\eta} \eta^{ab} G_{PN}\right)=0.
\end{array} \eeq Note that the derivative in the final term
acts on $\eta^{ab}$ depending on the choice of $\Sigma$.

The open string boundary condition applied at the string end
points $\sigma=\sigma_*$ is \beq  e^{\phi/2} \sqrt{-\eta}
\eta^{\sigma b} G_{MN}
\partial_b X^N = 0. \eeq
However, if the string is attached to a D-brane localized at $x^{M}=c^M$, then this
condition is replaced in the directions transverse to the brane
by the Dirichlet condition $X^M=c^M$.

There are also world sheet constraints from variation with respect to
$\eta_{ab}$ are \beq \label{con} \partial_a X^M \partial_b X^N
G_{MN}={1 \over 2} \eta_{ab}\eta^{cd} G_{MN}  \partial_c X^M
\partial_d X^N,\eeq which, with the form of $\eta_{ab}$ above become
\beq \label{con1}\dot{X}\cdot X' = 0,\eeq \beq
\label{con2}\dot{X}^2 + \Sigma^2 X^{'2} = 0, \eeq
where $\dot{X}$ indicates $\partial_{\tau}X$ and $X'$ denotes $\partial_{\sigma}X$.

We will also need to compute the energy of the strings that we
generate. That energy can be determined from the 10d space-time
stress energy tensor \beq \begin{array}{ccl} T^{MN} & = & -{2
\over \sqrt{-G} } { \delta S_P \over \delta G_{M N}(x)}\\&&\\ & =
& \frac{1}{2\pi\alpha'} \int d^2 \sigma e^{\phi/2}\sqrt{-\eta}
\eta^{ab} \partial_a X^M \partial_b X^N {\delta^{10}(x-X) \over
\sqrt{-G} }. \end{array} \eeq The conserved energy is then given
by the integral over the 9d space of $T^0_0$,
 \beq E = \frac{1}{2\pi\alpha'} \int d^2 \sigma
e^{\phi/2} \sqrt{-\eta} \eta^{ab} \partial_a X^0 \partial_b X_0
\delta(t-X^0). \eeq If we rewrite the delta function as \beq
\delta(t-X^0) = { \delta(\tau - \tau_*(\sigma)) \over |
\partial_\tau X^0(\tau_*(\sigma), \sigma)|},\eeq where
$\tau_*(\sigma)$ is the solution to
$X^0(\tau_*(\sigma),\sigma)=t$, we can then integrate over $\tau$
to find \beq \label{energy} E = \frac{1}{2\pi\alpha'} \int d
\sigma e^{\phi/2} \sqrt{-\eta} { \eta^{ab} \partial_a X^0
\partial_b X_0 \over | \partial_\tau X^0 |} |_{\tau =
\tau_*(\sigma)}. \eeq

\subsection{Separating Quark Solutions}
\label{sepQuark} In \cite{Chesler:2008wd} falling strings in AdS
with endpoints separating in the Minkowski directions $x^{\mu}$
were studied - those solutions describe states with massless
quarks in the basic ${\cal N}=2$ theory described by the D3-D7
system. For massless quarks, the D3 lies within the D7 and the
strings with ends on the D7 are free to come arbitrarily close to
the D3s at $r=0$. The solutions they found had the endpoints
separating in $x$ and, at late times, falling in $r$ along
geodesics. The centre of the string fell faster in $r$ leaving a
curved string between the endpoints.
They showed that kinks in the initial condition of the string were quickly
smoothed out by the inflation-like growth of the string.

We now wish to begin to make contact to QCD where, although the
current quark masses are small (for the light quarks), they
nevertheless obtain a large constituent mass from the chiral
symmetry breaking dynamics of the theory. For this reason, we will
work with the D3-D7 system in a configuration where the D3 and D7
do not coincide (i.e. the dual quarks are massive). Later, we will
consider the setup where we embed the D7 in the deformed AdS of Eq. (\ref{wallmetric}).

As shown in Figure 1, the motion of a string ending on the D7 is
generically a three dimensional problem. The string can move in
the Minkowski directions $x^{\mu}$ and the $r^{1-4}$ directions on
the D7 whilst the centre of the string may droop into the
remaining two $r^{5-6}$ directions of the geometry. For
simplicity, we will restrict the motion to two dimensions - that
is, we will look at strings lying at the point of closest approach
between the D3 and the D7, with ends separating in the $x^{\mu}$
directions. There is no hard computational block to performing the
more generic computation, but we believe we can pick out the
crucial physics within the simplified set-up.

So that our analysis can be compared to that in
\cite{Chesler:2008wd}, we will follow their example and define an
inverse radial coordinate \beq z^2 = {R^4 \over r_5^2 + r_6^2}. \eeq
The D7 brane embeddings are more complicated in these coordinates,
but
we will only work with strings at the point of closest approach to
the D3, $z_0$. The strings we study therefore lie at
$r_1^2+r_2^2+r_3^2+r_4^2=0$. The constituent mass of the quarks is
then $m_q = \frac{R^2}{2 \pi \alpha' z_0}$. The AdS boundary is
now located at $z=0$ and the IR (or the D3 branes) is at $z=
\infty$. The string then falls in the 3d space parameterized by
the Minkowski directions $x^0\equiv t,x^1\equiv x$ and radial
coordinate $z$ with metric \beq ds^2 = {R^2 \over z^2}
\left(-(dx^0)^2+(dx^1)^2 + dz^2\right). \eeq

The general initial conditions for such a string are parameterized
by specifying the embedding functions $t,x,$ and $z$ and their
derivatives at world sheet time $\tau=0$ as a function of
$\sigma$. As an example, consider the initial condition \beq
\label{ic1} t(\sigma,0) = 0, \hspace{0.5cm} x(\sigma,0) = 0,
\hspace{0.5cm} z(\sigma,0) = z_0. \eeq Note that this initial
condition corresponds to a configuration where the string is
initially contracted to a point.  Since $X'(\sigma,0)=0$, the
first world sheet constraint Eq. (\ref{con1}) is satisfied and the
second Eq. (\ref{con2}) is simply $\dot{X}^2=0$.

Of course, the point-like initial conditions Eq. (\ref{ic1}) are
not sufficient to specify the subsequent string evolution. We must
also specify the $\tau$-derivatives at $\tau=0$ consistent with
the constraint equations Eq. (\ref{con}). Consistent point-like
initial conditions are then specified by two free functions,
$\dot{x}(\sigma,0)$ and $\dot{z}(\sigma,0)$. The last derivative,
$\dot{t}(\sigma,0)$, is fixed by the constraints to be \beq
\dot{t}(\sigma,0)=\sqrt{\dot{x}(\sigma,0)^2+\dot{z}(\sigma,0)^2}.
\eeq As an example, we  consider strings with initial derivatives
\beq  \label{ic2} \dot{z}(\sigma,0)= C z_0 \sin \sigma,
\hspace{1cm} \dot{x}(\sigma,0)= D z_0 \cos \sigma, \eeq where the
string endpoints are located at $\sigma_*=0,\pi$. This set of
initial conditions has two free parameters $C$ and $D$ that
essentially set the string's momentum in the $z$ and $x$
directions respectively (the total string's $x$ momentum obviously
vanishes - $D$ sets the size of the momentum in the $x$ direction
of either half of the string). Strings generated from large $D$
and small $C$ will have end points that quickly separate in the
$x$ direction, so that the dual field theory state contains a
back-to-back quark/anti-quark pair with large momenta.

The string generated from these initial conditions has some energy.
Using Eq. (\ref{energy}), and the initial value of $\dot{t}(\sigma,0)$,
\beq
\dot{t}(\sigma,0) = z_0 \sqrt{C^2 \sin^2 \sigma + D^2 \cos^2
\sigma}, \eeq
we find that the total energy of this string is
\beq
\begin{array}{ccl} E & = & {1 \over 2\pi\alpha'} \int d \sigma g_{00}
{\partial_\tau X^0 \over \Sigma}|_{\tau=0} \\ &&\\
& = & {R^2 \over 2 \pi\alpha' z_0} \int d \sigma \sqrt{C^2 \sin^2 \sigma
+ D^2 \cos^2 \sigma}, \end{array} \eeq where we have normalised the
stretching function $\Sigma$ by $\Sigma(z_0)=1$. The constant $C$ lets us
choose the initial speed of the centre of the string in the $z$
direction whilst $D$ (in the large $D/C$ limit) picks the total energy of the string.

With these initial conditions we can solve the three
equations of motion (\ref{eom}) provided we also impose
the end point boundary conditions \beq t'(\sigma_*,\tau)=0, \hspace{0.4cm}
x'(\sigma_*,\tau) = 0, \hspace{0.4cm} z(\sigma_*,\tau) = z_0. \eeq
Note that our choices of initial time derivatives above are also
consistent with these boundary conditions. Finally, since the initial conditions satisfy the
constraint equations Eq. (\ref{con}), the evolved string satisfies them at all times.

A judicious choice of the stretching function allows NDSolve in
Mathematica to follow the evolution for a considerable time. Here
we pick \beq \label{sigads} \Sigma = z_0^2/z^2 \eeq To check the
consistency of the solutions along the time evolution we monitor
the total energy of the configuration using Eq. (\ref{energy}) -
for all our solutions energy is conserved at, at least, better
than the 1$\%$ level through the evolution. For the simple
solutions in AdS the conservation is much better than 1$\%$.

In Figure 3 we show the evolution of such a configuration with $C$
and $D=1$  and $z_0=0.5 R$. As the end points separate, the string
droops down into the bulk of $AdS$. Increasing $C$ only serves to push the middle of the string
down into AdS faster. Overall, this picture is consistent with
the usual expectations from static strings in AdS and with the
results described at zero mass in \cite{Chesler:2008wd}. In particular, the
fact that the string can penetrate indefinitely into the bulk of AdS reflects
the conformal $1/r$ potential between two
quarks. Of course, this behaviour is closer to the asymptotically
free regime of QCD than the confining phase. We now turn to
modelling the latter.

\section{Strings with a hard wall}

We will now study the motion of strings in the deformed-AdS
geometry Eq. (\ref{wallmetric}) to represent quark/anti-quark pair
production in a confining gauge theory. As for the strings we
studied in the last section, we will again set the string in
motion in the $x$ direction whilst localised in the $r^{1-4}$
directions at the point of closest approach of the D7 to the D3
brane - see Figure 2. The embedding function of the D7 brane will
not therefore play any role in the computation other than
determining that closest approach point, $z_0$. In practice, even
$z_0$ would be a phenomenological parameter that would need to be
fitted to the current quark mass - for the qualitative analysis
here we set $z_w/z_0 =2$ but the precise value is not important
(if one made the ratio very large then the quarks would become
heavy relative to the scale $\Lambda_{QCD}\sim 1/z_w$ representing
a large bare and constituent quark mass).

\begin{centering}
\includegraphics[width=70mm]{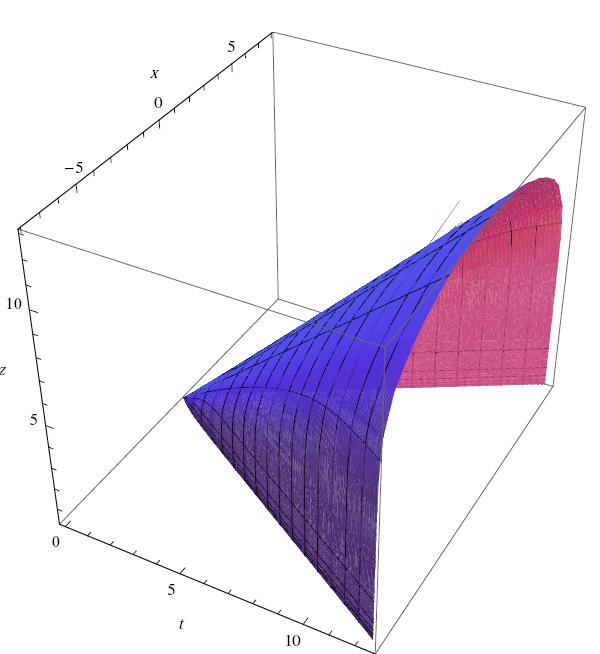} \\ \vspace{1cm}
\includegraphics[width=80mm]{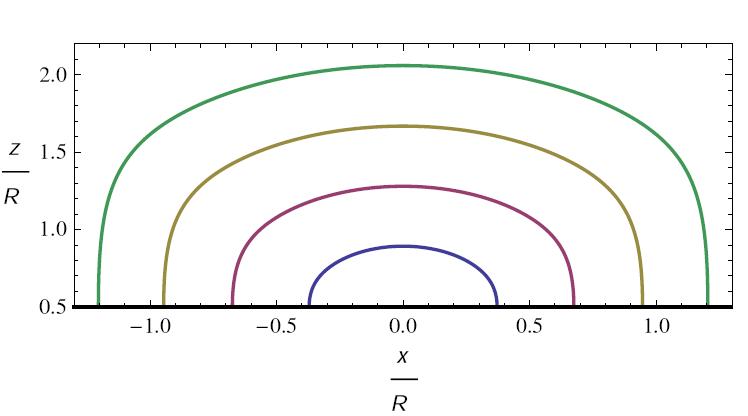}
\end{centering}
{\small{Figure 3: The evolution of a string with end points on a
D7 brane at $z_0=0.5 R$ in pure AdS with the initial conditions
Eq.  (\ref{ic1},\ref{ic2}) and parameters $C=D=1$. The top plot
shows the string worldsheet evolution and the bottom a series
constant time slice shots of the strings motion at
$t=0.4R,0.8R,1.2R,$ and $1.6 R$.}}
\par

We can now consider the evolution of strings in this geometry
using initial conditions like those we used above in pure AdS -
that is, we look at a point-like string with separating end
points. For simplicity, we will take almost exactly the same
initial conditions. We maintain the conditions in Eqs.
(\ref{ic1}),(\ref{ic2}) but to satisfy the constraint Eq.
(\ref{con2}) with the new metric, we need a new initial condition
for $\dot{t}(\sigma,0)$:

\beq \dot{t}(\sigma,0) = z_0\sqrt{(C^2/A^2(z_0)) \sin^2 \sigma +
D^2 \cos^2 \sigma}. \eeq Since $A(z)$ rapidly becomes unity away
from the wall $z=z_w$, these strings are generated from
essentially the same initial conditions as in the pure AdS
analysis. We use the same stretching function $\Sigma$ as in the
AdS case here, Eq. (\ref{sigads}). We plot the resulting string
evolution in Figure 4.

\newpage

\begin{centering}
\includegraphics[width=80mm]{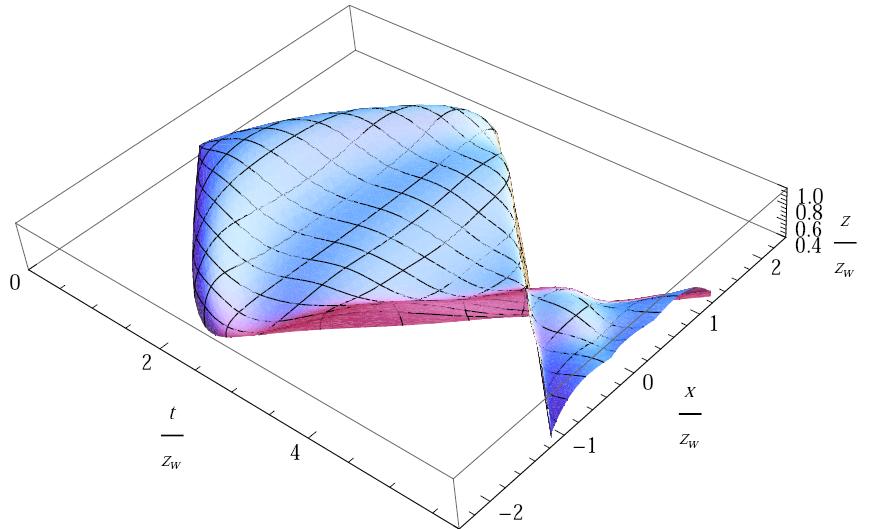}
\\ \vspace{1cm}
\includegraphics[width=80mm]{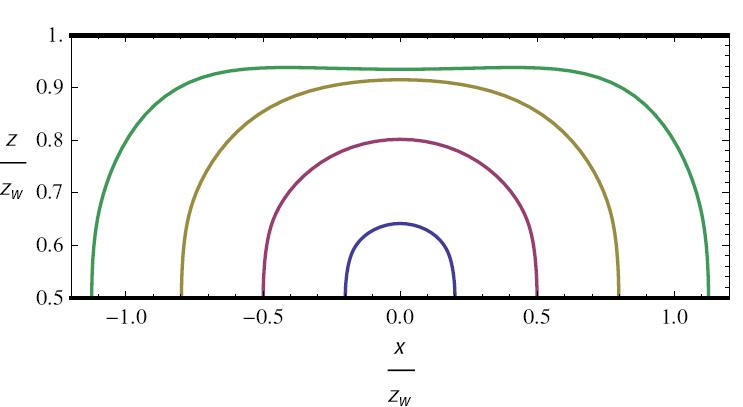}
\end{centering}
{\small{Figure 4: String evolution in a hardwall geometry. Here we
set $C=2.5, ~D=0.5$ in the initial conditions Eqs.
(\ref{ic1},\ref{ic2}), and choose $z_0=0.5 z_w$ to set the quark
mass. The top plot shows the world sheet evolution. The bottom
plot shows time slices through the world sheet at times
$t=0.2z_w,0.5z_w,0.8z_w,$ and $1.1z_w$. }}
\par

The solution plotted in Figure 4  shows many of the properties one
would expect. In particular, as the centre of the string
approaches the hard wall, it is repelled and a straight section of
string builds against the wall - this is the formation of the
naive QCD string. We are working at infinite $N$ where string
breaking is forbidden and so, rather than breaking, the string
continues to evolve. The centre of the string begins to bounce off
the hard wall and starts oscillating between the position of the
D7 brane and the hard wall. Meanwhile, as the potential energy in
the string grows and the quarks separate, the quarks' kinetic
energy begins to be sucked into the string - they slow. Eventually
the quarks are brought to a halt and the string between them
begins to contract, reversing the quarks' motion until they pass
through each other. This oscillation will continue indefinitely in
the absence of string breaking.

We conclude that the inclusion of a hard wall does indeed begin to
move the description of quark pair production closer to the
expected behaviour in QCD. In the next section we will discuss
how to include string breaking and thereby allow two separated jets to emerge.

\begin{centering}
\includegraphics[width=80mm]{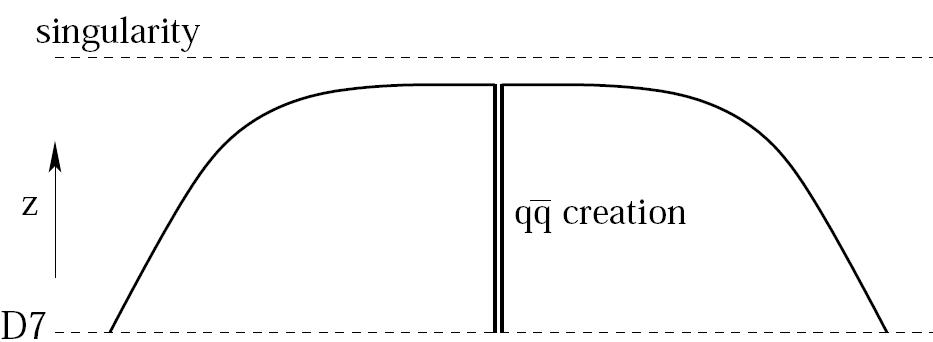}
\end{centering}
{\small{Figure 5: Sketch of the insertion of a quark anti-quark
pair into the string evolution to represent string breaking. }}
\par

\section{String Breaking}

Our analysis of hadronization so far lacks one crucial ingredient
with respect to QCD - there is no string breaking and therefore
no hadronization! String breaking is an inherently $1/N$ effect
and so absent from the AdS/CFT Correspondence in the tractable
limit we study. Some progress has been made in studying this process
though - if the string lies within the D7 worldvolume where it can 
break directly the rate is computable \cite{break1, break2, break3,
break4} and is of order $1/N$. When the string lies away from the D7 
there must be a quantum fluctuation that brings the string back to
the D7 where it can break - estimates for this process can be found in 
\cite{break3, break4}. Here we do not wish to add to these
computations.
Our instinct instead is to assume that in QCD the string will break
with essentially probability one as soon as there is of order
$\Lambda_{QCD}$ of potential energy stored in the string. We can
model this on the gravity side by breaking our string at the
corresponding time.

We can insert a quark/anti-quark pair into the evolution of our
strings by hand. An unpaired quark corresponds to a string
stretched from the D7 brane to the D3 branes. A static solution of
this form with a new choice of stretching function ($\Sigma =
A(z)/A(z_0)$) is \beq \begin{array}{c} t(\tau,\sigma) = {{2 \left(
z_0 - z_s \right)}\over {\pi A(z_0)}}  \tau \\ \\ z(\tau, \sigma)
= {{2 \over{\pi}}\left(z_0-z_s\right)\sigma+2z_s-z_0},
\end{array} \eeq where $x$ is constant. This solution satisfies the equations of motion, the
constraint equations and the boundary conditions for all $\tau$
and $\sigma$. It is a straight string stretching from $z_0$ (at
$\sigma=\pi$) through the  point $z_s$ (at $\sigma=\pi/2$).

We can therefore split our string in the centre and complete it to
the D7 brane with a straight string segment as shown in Figure 5.
As an example we will split the string solution shown in Figure 4
at the point where the centre of the string starts to bounce off
the wall ($t=1.13z_w$) - the QCD string has roughly just formed at this
point in the time evolution. At the splitting time $t_s$ ($\tau_s$
in world sheet time) when $t(\tau_s, \pi/2)= t_s$ and $z(\tau_s,
\pi/2) = z_s$ we use the new initial conditions for the range
$\pi/2 < \sigma \leq \pi$.

Strictly inserting the extra string length does not conserve
energy for the half string piece - our solutions below do have
sufficient initial energy that the straight string is slightly
less than a 10$\%$ correction to the total energy. We don't expect
the non-conservation to have any great effect on the qualitative
behaviour of the solutions we display. In fact since we only study
the evolution of one half of the initial string configuration one
could imagine that some asymmetry in the distribution of energy
between the two halves might be induced in the string breaking in
any case. One could also splice out a straight string section in
the middle of the initial string before inserting the vertical
string pieces back to the D7 brane but we again would expect to
see little qualitative change in the behaviour of the solutions.

The time evolution of the half string segment can now be followed
and we show such a numerical solution in Figure 6. The solution
again behaves in accordance with naive expectations - the fast
moving end point of the string continues to move, whilst two kinks
in the string induced by the breaking propagate to the end points.
When the kink arrives at the static end point, it is jerked (along
with the entire string segment) in the direction of motion of the
fast end point. Similar evolution of broken strings
may be found in \cite{Peeters:2004pt}.
Following the evolution without further string
breaking leads to oscillations of the end points in the segment's
centre of mass frame, which separates infinitely far away from the
other broken string segment. We also plot in Figure 7 the motion
of the two end points of the string through this evolution - the
string end points are special because they source a gauge field on
the D7 brane world volume which corresponds to rho meson
production. We will compute this production in the next section.
\vspace{2cm}

\begin{centering}
\includegraphics[width=80mm]{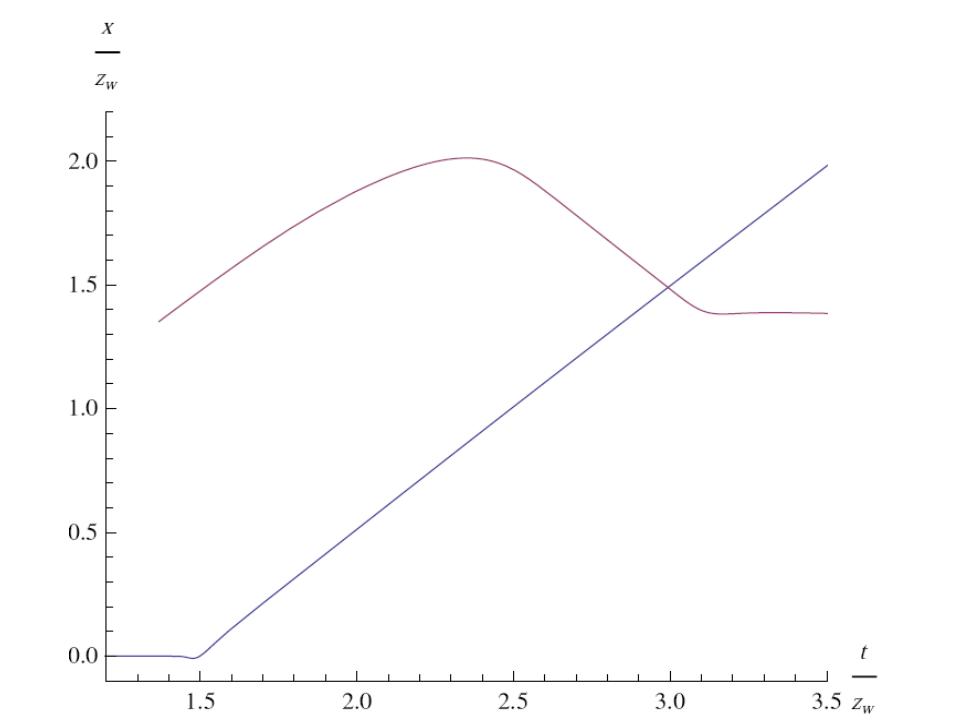}
\end{centering}
{\small{Figure 7: Plot of the string end point motion for the
configuration in Figure 6. }}
\par

\begin{centering}
\includegraphics[width=80mm]{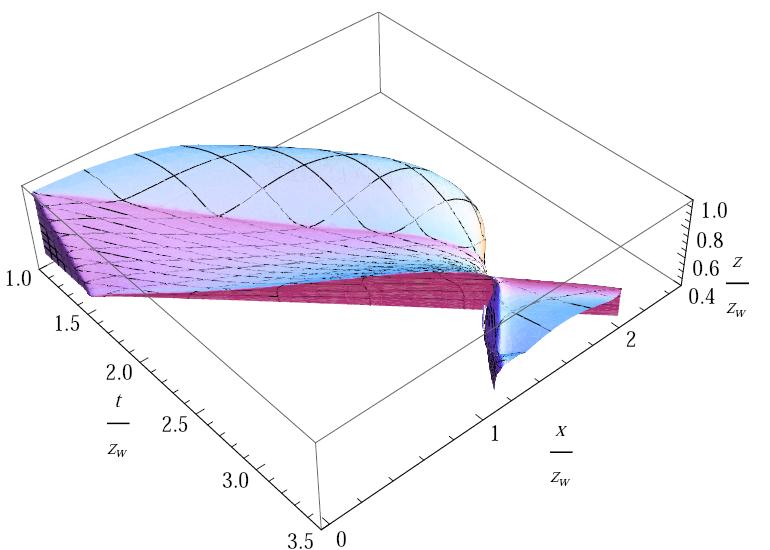}
\\ \vspace{1cm}
\includegraphics[width=80mm]{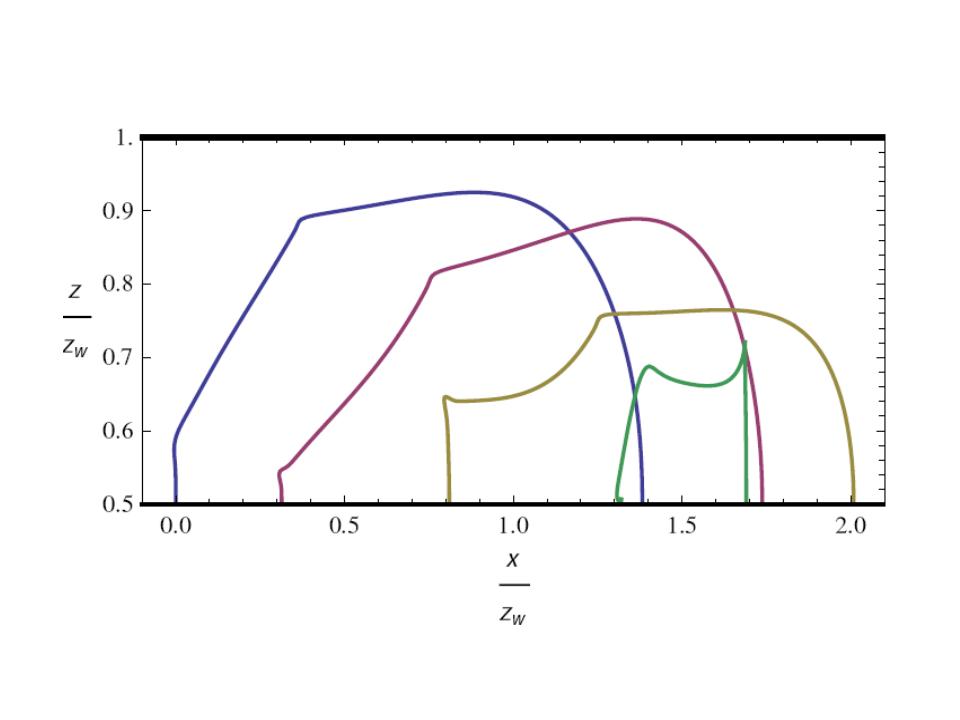}
\end{centering}
{\small{Figure 6: We show a string world sheet evolution after
imposing string breaking. The initial condition involves half of
the string from Figure 4 at time $t=1.13z_w$ broken and extended by a
straight string back to the D7 brane at $\sigma=\pi/2$ to
$\sigma=\pi$. The top plot is again the world sheet evolution and
the bottom plot slices taken at $t=1.4z_w,1.8z_w,2.3z_w,$ and $3.2z_w$}}
\par
\newpage

\section{Rho meson production}

To understand how the string solutions above radiate energy into
hadronic modes, one must study the electromagnetic theory on the
surface of the D7 brane. For simplicity, we will assume that the
D7 of Figure 2 has been pushed out sufficiently far away from the
hard wall in the geometry Eq. (\ref{wallmetric}) that we may treat
the 10d geometry as AdS. Further, we assume that the D7 is
essentially
flat. That is, we will approximate the D7 embedding with the flat
embedding  of a D7 in pure AdS down to $r=r_0$ with induced metric
\beq P[G]\equiv g = {\rho^2+ r_0^2 \over R^2} dx_4^2 + {R^2\over
\rho^2 + r_0^2}(d\rho^2 + \rho^2 d \Omega_3^2). \eeq Here, $\rho$
is the radial direction on the world volume of the D7 so that
$\rho^2= r_1^2+r_2^2+r_3^2+r_4^2$. Although these approximations
may seem a little crude we will see that the $\rho$ dependence of
the problem enters essentially just through the mass of the
mesonic states which can therefore be easily altered to any more
complicated set up.

The end points of the string act as electrically charged sources
for the gauge field that lives on the world volume of the D7
brane. The equation of motion for that gauge field follows from
the variation of the electromagnetic action on the brane, \beq
S_{\rm EM}= -\frac{1}{4}\int d^8x \sqrt{-g}F_{ab}F^{ab}+\int d^8x
\sqrt{-g}j^{a}A_a. \eeq The variation of the gauge field gives
both the equation of motion and the boundary action, \beq
\label{varyS} \delta S_{\rm EM} = \int d^8 x \sqrt{-g} \delta A_b
\left[ \frac{1}{\sqrt{-g}}\partial_a(\sqrt{-g}F^{ab})+j^b
\right]+\delta S_{\rm bdy}, \eeq so that the equations of motion
are just Maxwell's equations, \beq \label{eqom} {1 \over
\sqrt{-g}} \partial_a ( \sqrt{-g} F^{ba}) = j^b, \eeq and there is
a boundary action (at $\rho= 1/ \epsilon \rightarrow \infty$) \beq
\delta S_{\rm bdy} =
-\int_{\rho=\frac{1}{\epsilon}}d^4xd\Omega_3\sqrt{-g} \delta
A_aF^{\rho a}. \eeq Let us first understand the solutions in the
absence of a source. \vspace{3cm}

\subsection{Rho mesons}

The propagating modes of the gauge field arrange themselves into
multiplets of the $SO(4)$ isometry group of the $S^3$. The
resulting Kaluza-Klein fields each map to different operators of
the gauge theory; in particular, the singlet on the $S^3$ maps to
a conserved baryon current \cite{Mateos}. To study this current,
we therefore give the solutions of Eq. (\ref{eqom}) without
sources for the modes that only have non-zero $A_{\mu}$ and are
singlets on the $S^3$. We impose the gauge choice $\nabla_\mu
A^\mu = 0$. The equation of motion in the absence of sources is
then \cite{Mateos} \beq \label{rhoEq}\mathcal{D}A_{\mu}=-{1 \over
\rho^3}
\partial_\rho (\rho^3 \partial_\rho A_\mu) - { R^4 \over
(\rho^2 + r_0^2)^2} \nabla_{(4)}^2 A_\mu  = 0, \eeq where
$\nabla_{(4)}^2=\eta^{\mu\nu}\partial_{\mu}\partial_{\nu}$ is the
scalar Laplacian in Minkowski space and $\mathcal{D}$ is a
second-order differential operator. Fourier transforming in the
Minkowski directions, we can write the equation above as an
eigenvalue equation in the radial coordinate with eigenfunctions
$f_n(\rho)$ given by \beq \label{basis} f_n(\rho) = I_n {_2F_1(-n,
-n-1,2; -\rho^2/r_0^2) \over (\rho^2+r_0^2)^{n+1}},\eeq where the
$I_n$ are normalization constants, $n=0,1,2...$, and eigenvalues
\beq \label{massmes} M_n^2 = 4(n+1)(n+2) {r_0^2 \over R^4} \eeq
The solutions to Eq. (\ref{rhoEq}) are therefore  given by the
modes $A_{\mu,n}=\epsilon_{\mu}f_n(\rho)e^{ik_n\cdot x}$ with
$k_n^2=-M_n^2$.

These states have a discrete mass spectrum and are identified with
the rho mesons of the dual gauge theory. The factor of $r_0^2/R^4
\sim m_q^2/\lambda$ indicates that the meson masses are much
smaller than the quark mass at large 't Hooft coupling. Moreover,
the $f_n$ (with appropriate choices of the $I_n$ normalizations)
are orthonormal functions (subject to the weight factor $w$) \beq
\int d\rho ~ w ~f_n f_m = \delta_{mn}, ~~~ w = {\rho^3 R^4 \over
(\rho^2+r_0^2)^2}. \eeq The corresponding choice of normalization
is given by \beq \label{In} I_n^2=2(n+1)(n+2)(3+2n). \eeq We are
using the solutions appropriate for the ${\cal N}=2$ D3-D7
configuration here rather than those for a bent brane as in Figure
2. However, as we will see, the holographic directions only enter
into our final radiation computation through the masses they endow
the four dimensional rho mesons and the value of $I_n$, the
normalization of the wave functions. In the more complicated case,
one could simply switch the spectrum and normalizations as
appropriate.

\subsection{Green's Functions}

To observe  the emission of rho mesons by the string end points we
will solve (\ref{eqom}) by means of a Green's function for the
field $A_{\mu}$. For the minimum-energy configuration quark (a
string connecting the D3 and D7 at their point of closest
approach) the endpoint lies at $\rho=0$  where the volume of the
three-sphere is zero and hence the Green's function is a constant
on the three-sphere (the equivalent of the $f_n$ functions for
R-charged states fall to zero at $\rho=0$). This implies that
there is no production of $R$-charged rho mesons associated with
non-trivial spherical harmonics on the $S^3$. For a more generic
string motion such states would be produced. Moreover, since the
sources do not move in the $r^{1-4}$ four-plane, both the radial
and angular components of the source current $j^a$ vanish and thus
$A_{\rho}$ and $A_i$ (the components of the gauge field along the
$S^3$) also vanish in this gauge. We can therefore consider only
the Minkowski components of the Green's function.

Having chosen the Lorentz gauge, that Green's function satisfies
\begin{equation} \label{Geqn}
\mathcal{D}G^{\mu'}_{\; \mu} = \frac{1}{\rho^3}
\delta^{\mu'}_{\mu} \delta(\rho-\rho') \delta(x^{\nu}-x^{\nu'}),
\end{equation}
where $\mathcal{D}$ is the differential operator defined in Eq.
(\ref{rhoEq}). Since the equation of motion for the gauge field in
the presence of our source is given by \beq \mathcal{D}A_{\mu} =
\eta_{\mu\nu}j^{\nu}, \eeq the full solution for an arbitrary
current distribution $j^{\mu}$ follows from the convolution
integral
\begin{equation}
A_{\mu}(x) = \int d^8x' \sqrt{-g} \;
G^{\mu'}_{\;\mu}(x,x')\eta_{\mu'\nu'} j^{\nu'}(x').
\end{equation}

The actual current distribution will be localized on the worldline
of the string endpoint and will take the form $j^{\mu}=q
\dot{x}^{\mu} \delta^8(x)$ where the dot represents
differentiation with respect to proper time.

In order to obtain the Green's function, let us expand in the
basis of eigenfunctions describing the rho mesons used in Eq.
(\ref{basis}) so that $G^{\mu'}_{\;
\mu}(\rho,x^{\nu};\rho',x'^{\nu}) = \sum_n f_n(\rho)f_n(\rho')
\bar{G}^{\mu'}_{n, \mu}(x^{\nu},x'^{\nu})$. Inserting this form
into Eq.(\ref{Geqn}), multiplying by $\rho^3 f_m$ and integrating
over all space we find that the four-dimensional functions
$\bar{G}_n$ are just the Green's functions for massive vectors in
Minkowski spacetime with masses corresponding to the rho meson
masses.

\subsection{Boundary Data}

The near-boundary behaviour of the gauge field is related to the
one-point function of the dual conserved baryon current in the
field theory. In particular, that one-point function is given as
\beq \left\langle J^{\mu}(x^{\nu})\right\rangle =
\lim_{\epsilon\rightarrow  0}\frac{\delta S_{\rm SUGRA}}{\delta
A_{\mu}(x^{\nu},1/\epsilon ) }, \eeq where $S_{\rm SUGRA}$ is the
on-shell bulk gravity action and the bulk gauge field $A_{\mu}$ is
the singlet mode on the $S^3$. Using the variation of the bulk
action in Eq. (\ref{varyS}), the boundary current is simply \beq
\label{bdyJ} \left\langle J^{\mu}(x^{\nu})\right\rangle =
-\lim_{\epsilon\rightarrow 0}
\rho^3\eta^{\mu\nu}\partial_{\rho}A_{\nu}(x^{\nu},\rho)|_{\rho =
1/\epsilon}. \eeq We can therefore write a bulk-to-boundary
Green's function that relates the bulk source to the boundary
current. In particular, we write \beq \left\langle
J^{\mu}(x^{\nu})\right\rangle = \int d^8 x'
\sqrt{-g}\mathcal{G}^{\mu}_{\;\mu'}(x^{\nu};x')j^{\mu'}(x'), \eeq
where we define the bulk-to-boundary Green's function
$\mathcal{G}$ as \beq \mathcal{G}^{\mu}_{\;
\mu'}(x^{\nu};x'^{\nu},\rho')\equiv  \sum_n 2(-1)^{n}I_n
f_n(\rho')\bar{G}^{\mu}_{n, \mu'}(x^{\nu},x'^{\nu}), \eeq where
the $f_n$ are the eigenfunctions in Eq. (\ref{basis}) and
$\bar{G}_n$ is the 4d Green's function for a massive vector as
before. The factor of $2 (-1)^n I_n$ comes from the insertion of
the near-boundary expansion of the $f_n$'s, \beq
f_n(\rho)=(-1)^{(n+1)}I_n \frac{1}{\rho^2}+O(\rho)^{-3}, \eeq into
the form of the boundary current in Eq. (\ref{bdyJ}).

\subsection{Retarded Potential}

We have now reduced the problem to solving for each mode $G_n$ the
retarded potential for a massive field in flat space. The retarded
potential takes the form \cite{Poisson:2003nc}
\begin{equation}
\bar{G}^{\mu}_{\; \mu'} = {1 \over 4 \pi} \theta(t-t') \left (
\delta(\sigma) + V(\sigma) \theta(-\sigma) \right )
\delta^{\mu}_{\; \mu'}.
\end{equation}
Here we use the Synge world-function $\sigma = \frac{1}{2}
\eta_{\mu \nu}(x-x')^{\mu} (x-x')^{\nu}$.  The non-singular part
of the solution is  given by $V(\sigma)= - \frac{M_n}{\sqrt{-2
\sigma}} J_1(M_n \sqrt{-2 \sigma})$ where $J_1$ is the Bessel
function of order $1$.

\subsection{Static String End Point}

As a first example of using this formalism we will compute the
baryon density around a static quark. Consider such a charge at
$x=0$ and at $\rho=0$ ($r=r_0$), the point of closest approach on
the D7 brane. We will concentrate on the temporal component of the
gauge field $A^0$ which is dual to the operator $\bar{\psi}
\gamma^0 \psi$, the quark density.

One seeks to evaluate the integral over the past trajectory of a
point source moving with a constant speed in a `static gauge'
given by $x'= \beta t'$.  Doing the spatial integral using the
fact that the source is located at a point in the space-like
dimensions leaves one with the integral
\begin{equation} \begin{array}{l}
\left\langle J^0(x^{\nu})\right\rangle_n =  \frac{2(-1)^n I_n^2
q}{4 \pi}  \times
\\ \\ \left. \right. \hspace{0.5cm} \int \; dt' \; \theta(t-t')
\left ( \delta(\sigma) + V(\sigma) \theta(-\sigma) \right )
\frac{d \tau}{dt'} \cdot \frac{d t'}{d\tau} \end{array}
\end{equation} where we have used the fact that $f_n(\rho=0)=I_n$.

The time component of the four-velocity is actually cancelled by a
Lorentz factor coming from the splitting of Minkowski spacetime
into space-like sections when integrating along the particle
worldline.  The two contributions to the integral are easily
computed (for the non-singular piece due to the massive field
using the integration variable $u \equiv M_n \sqrt{-2 \sigma}$))
giving the correctly Lorentz-covariant expression (the $\gamma$ is
the usual boost factor)
\begin{equation}
\left\langle J^0(x^{\nu})\right\rangle_n= \frac{2 (-1)^n I_n^2
q}{4 \pi} \frac{\gamma \; e^{-M_n \left ( \sqrt{ \gamma^2 (x-\beta
t)^2+y^2+z^2} \right )}}{\sqrt{\gamma^2 (x- \beta t)^2+ y^2+z^2}}
\end{equation}

In the rest frame of the point source this reduces to the usual
Yukawa form.  The full solution is a sum over modes weighted by
the $f_n$ normalizations $I_n^2$, see Eq. (\ref{In}). There is a
rapid rise in these normalizing factors with $n$ which is due to
the end point of the string being a delta function. Away from
$N\rightarrow \infty$ one would expect the string to have some
width and the expansion to truncate at some intermediate $n$. In
any case this rise is not faster than the exponential fall off of
the solutions so the physics away from the source is still
dominated by the lightest modes. The Green's function converges
for all $|x|>0$ due to the exponential factor in the Yukawa
potential of each partial wave. The behaviour is dominated by the
lighter modes at distances comparable to the Compton wavelength of
the lightest mode.  We interpret this Green's function as the
`dressing' of an isolated quark by a cloud of mesons.  Holography
gives the relative amounts of each of the excited states in the
cloud.

\subsection{Radiation From String End Points}

We now have a framework in which the emission of mesons  can be
modelled using the techniques of classical relativistic wave
equations.  The retarded Green's function is straightforwardly
integrated over the past worldline of an accelerating endpoint to
give eg. for a particle moving in the $x$-direction (the $u$
variable is as defined in the preceding section)
\begin{equation}
\begin{array}{l} \label{retard}
\left\langle J^0(x^{\mu})\right\rangle _n= \frac{2(-1)^n I_n^2
q}{4 \pi} \left[ \frac{1}{t-t'(\sigma=0)- \frac{dx'}{dt'} \left (
x-x'(\sigma=0) \right )}+ \right.\\ \\~~~~ \left. \hspace{2cm}
\int_0^{\infty} \; du \frac{J_1(u)}{t-t'-\frac{dx'}{dt'}(x-x')}
\right]
\end{array}
\end{equation}

We will again plot the baryon number density which is
holographically encoded by the sum over the $J_n^0$.  It may be
noted that the plots we obtain give the superposition of radiated
baryon density and the static baryon density associated with the
probe quark.  An elementary prescription is available for
computing the reaction force on the probe quark due to the
radiation (by differencing the advanced and retarded potentials)
but this is not what we are interested in here (it involves a
negative counting of the non-causal advanced potential and so
would not produce a plot resembling meson {\it emission}).  In
holographic scenarios (large $N$) the force exerted on the quark
by the dynamics of the colour flux tube far exceeds the reaction
force from meson emission anyway.  In the case of an {\it
instantaneous} acceleration it is possible to subtract the
appropriate static and boosted solutions inside and outside of the
particle's light cone but we do not apply this here.

\subsubsection{Massless Meson Limit}
In the strict $\lambda \rightarrow \infty$ limit the meson masses
are very small relative to the string mass (see (\ref{massmes}).
At least for the lightest members of the tower, it is therefore
interesting to compute the radiation into a massless gauge field
on the D7. For this case $\sigma = {1 \over 2} \left(- (t-t')^2 +
(x-x'(t'))^2 \right)$. The first term in (\ref{retard}) then gives
\beq \label{massless} \begin{array}{ccl} \left\langle
J^0(x^{\nu})\right\rangle_n & = & \frac{2(-1)^n I_n^2 q}{4 \pi}
\int^t_{-\infty} dt'
\delta (\sigma)\\ &&\\
& =& \frac{2(-1)^n I_n^2 q}{4 \pi} \int {d \sigma \delta (\sigma)
\over \left((t'-t) +
(x'-x) { d x' \over d t'} \right) } \\ &&\\
&=& \frac{2(-1)^n I_n^2}{4 \pi} \frac{1}{t-t'(\sigma=0)- \dot{x}_0
\left ( x-x'(\sigma=0) \right )}\end{array} \eeq This is
straightforward to evaluate for acceleration kicks such as those
we found for the string end points in Figure 7 above. For example
the static end point is accelerated quickly to a constant speed.
As an example form for the function $x'(t')$ that describes a
stationary particle accelerating to a final speed $a$ we take \beq
\label{kick} x'(t') = {a b \over \pi} + a t' \left( {1 \over 2} +
{1 \over \pi} \tan^{-1} \left ({t' \over b}\right ) \right)\eeq
$b$ here controls the time interval over which the acceleration
occurs - we plot some sample trajectories in Figure 8.
\vspace{1cm}

\begin{centering}
\includegraphics[width=80mm]{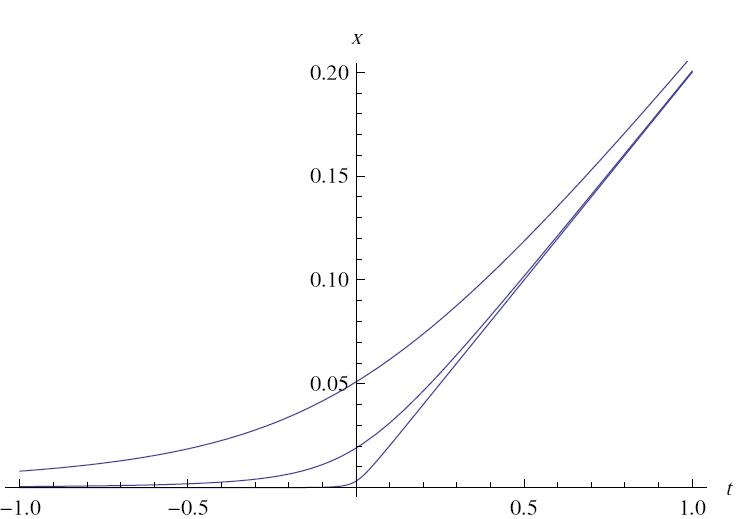}
\end{centering}
{\small{Figure 8: Plots of the function $x'(t')$ in (\ref{kick})
used to parameterize the motion of an accelerating point source.
The parameter $a$ controls the final speed and is set to $a=0.2$
here. $b$ controls the time scale of the acceleration and the
plots show $b=0.8$ (top), $b=0.3$ (middle) and $b=0.05$
(bottom).}}
\par \vspace{1cm}

It is a simple matter to plot the resulting wave induced. Emission
is typically a spherical shell radiating from the point of
acceleration - there is an SO(2) symmetry in the $y,z$ coordinates
so we shall plot the intensity of the wave in the $x,y$ plane at
$z=0$. Examples of the gauge field produced are shown in Figure 9.
The radiative piece is visible along with the `hill' of the
boosted static potential. A clear, narrow emission wave is
observable. For larger values of the final speed $a$ the forward
emission is typically enhanced relative to the backwards emission,
and the overall emission is greater. For smaller acceleration
times (smaller $b$) the wave front simply becomes narrower. For
the accelerations of the string end points in Figure 7 we expect
precisely such emission of the lower mass members of the mesonic
tower.

\begin{centering}
\includegraphics[width=80mm]{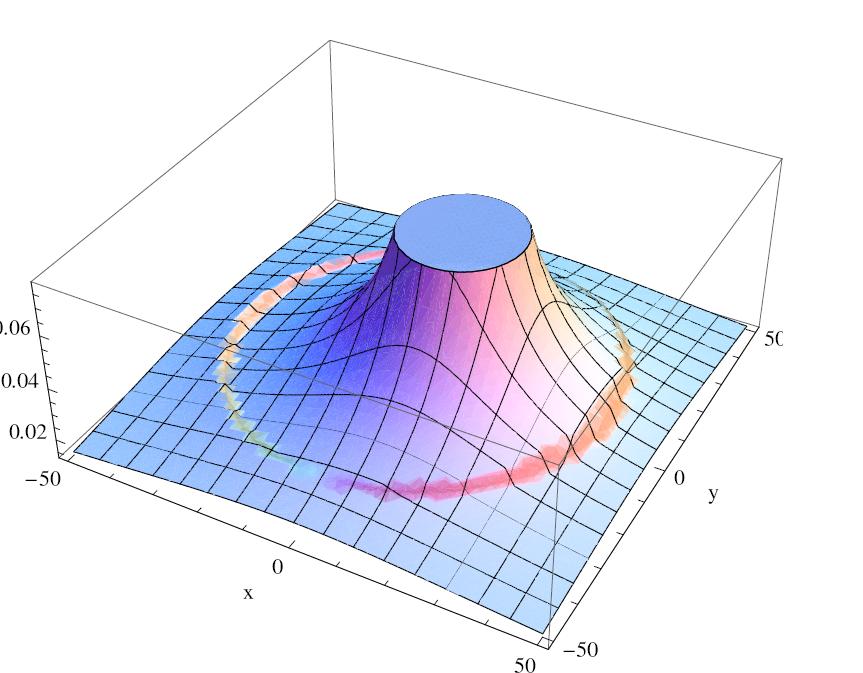}
\par
\includegraphics[width=80mm]{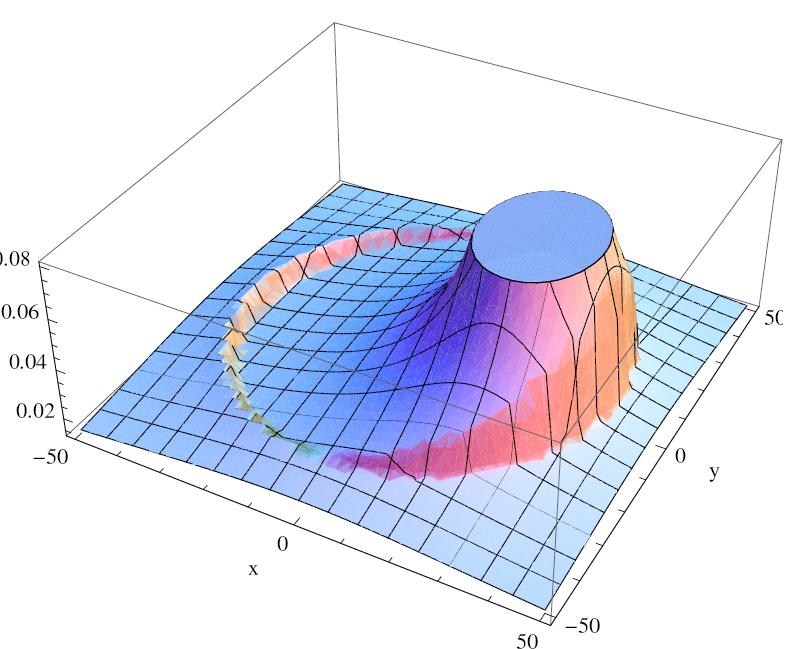}
\end{centering}
{\small{Figure 9: The radiation of light mesons by a quark  given
an impulse in the positive $x$-direction, shown in the $z=0$
plane.  The plot shows the density of the emitted (radiative part
of field) and bound (boosted static part) mesons.  The top plot is
for a terminal velocity of $0.2 c$ and below is $0.6 c$ (in both
plots the parameter $b=0.2$).}}
\par

\subsubsection{Massive Meson Limit}
For members of the meson tower with masses close to the quark mass
(very high $n$ at large 'tHooft coupling) we must compute the
non-singular term in (\ref{retard}) which involves numerical
integration of the Bessel function.  In Figures 10 and 11 we show
the effect of increased meson mass on the radiated mesons (the
meson mass should be compared to the inverse time over which the
string end point is accelerated). As the meson mass is increased
we find emission of the more massive states are suppressed.

In the massive case the waves are dispersive and produce an
interesting pattern which is not just a wave localized on the
light-front. The `wavy' emission of meson density can be
considered to be arising from quantum-mechanical interference
effects.

In principle we could sum over the emission of all of the meson
states. At large 'tHooft coupling though there are many states
lighter than the quark mass so the result would be unilluminating.
The precise form of the meson masses and the coefficients $I_n^2$
are also model dependent. However, we believe that the
computations we have made show how in principle the radiation
could be computed and give a good understanding of the generic
features of that meson radiation. \vspace{4cm}


\begin{centering}
\includegraphics[width=35mm]{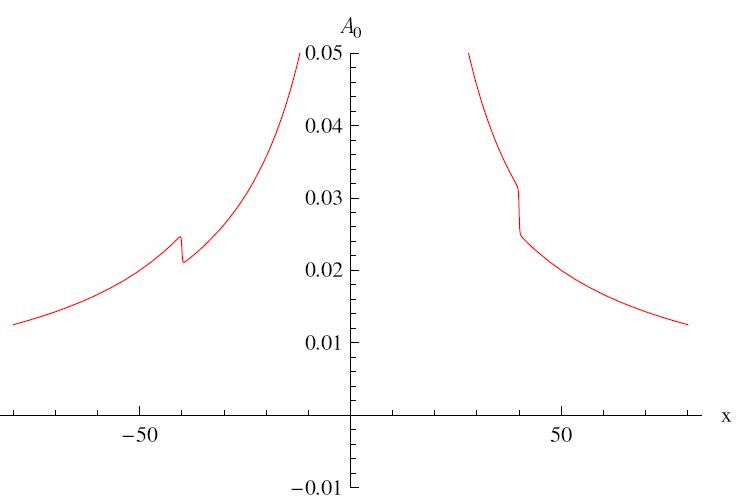}
\includegraphics[width=35mm]{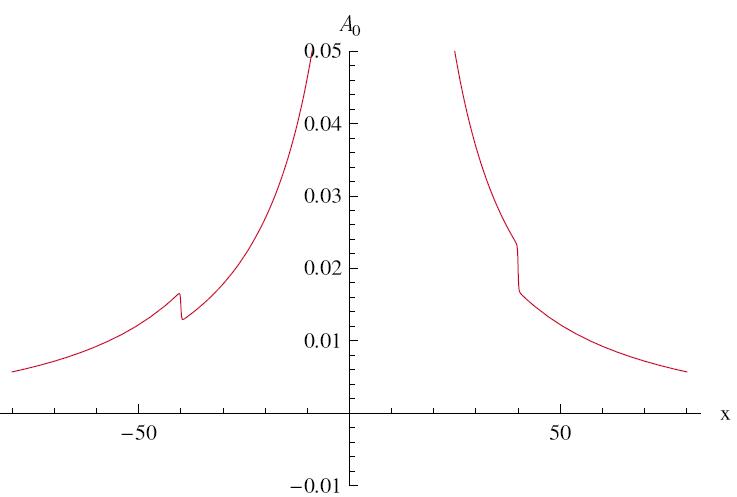}
\par
\includegraphics[width=35mm]{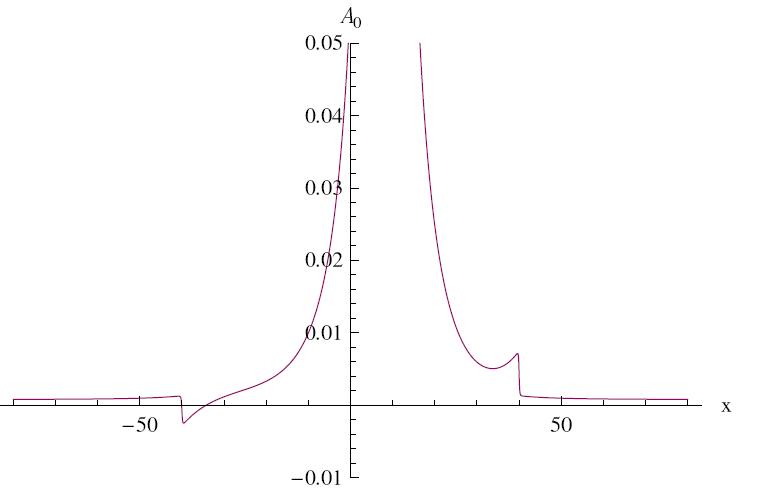}
\includegraphics[width=35mm]{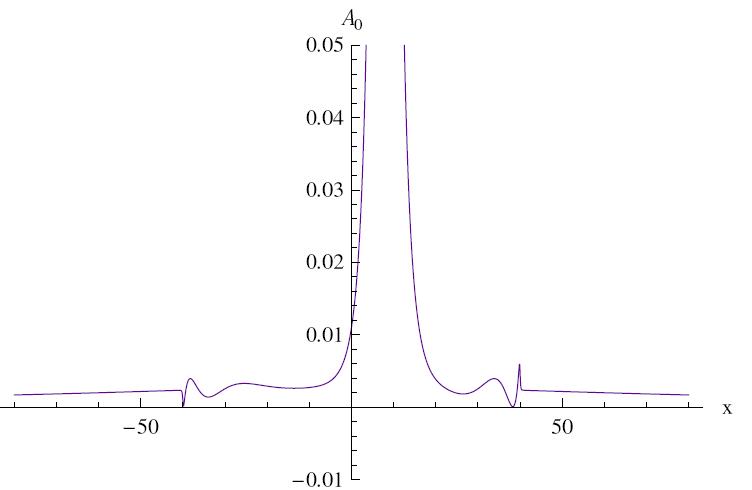}
\par
\centerline{\includegraphics[width=35mm]{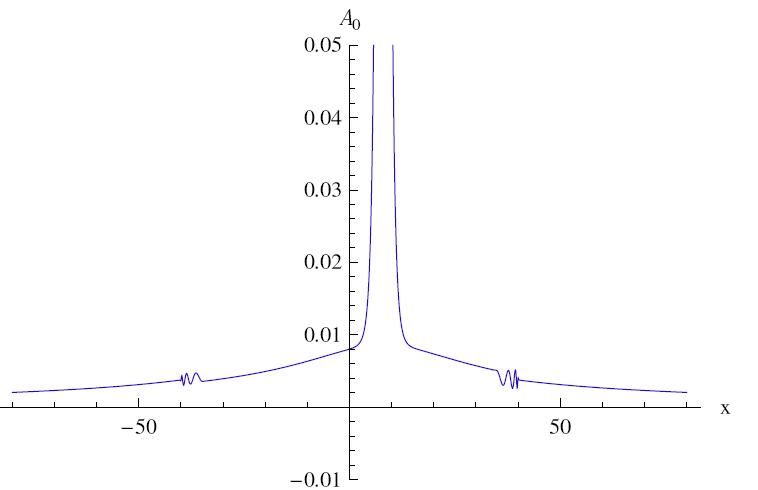}}
\end{centering}

{\small{Figure 10: The radiation of massive mesons by a  quark
given an impulse in the positive $x$-direction, plotted along the
$x$-axis.  The plot shows the density of the emitted mesons.  The
meson masses increase through the plots as
$0$,$10^{-2},10^{-1},\frac{1}{3},1$. Note the background static
field peak becomes narrower as the mass is increased.}}
\par

\begin{centering}
\includegraphics[width=80mm]{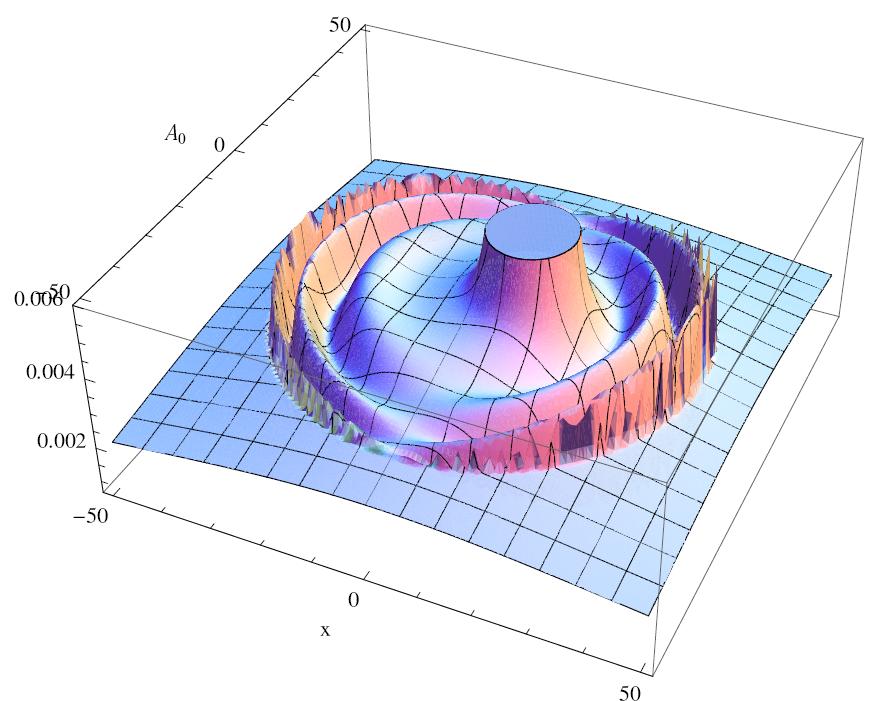}
\end{centering}
{\small{Figure 11: Emission of massive vector mesons
($m=\frac{1}{2}$).   The parameters $a=b=0.2$. }}
\par

\section{Lessons for QCD}

We have explored how a hadronization event happens in a gauge
theory that has the degrees of freedom of ${\cal N}=4$ super
Yang-Mills in the UV but a deformation that leads to a back
reacted hard wall and confinement in the IR. This theory, of
course, is not QCD but the generic picture that emerges may have
some lessons for the construction of a phenomenological model of
hadronization in QCD.

In particular we have suggested a picture in which the initial
quark anti-quark pair separate, growing a string between them that
dips into a holographic radial direction. Initially the string's
energy represents a $1/r$ potential between the quarks reminiscent
of the asymptotically free regime of QCD. In actuality the ${\cal
N}=4$ theory is strongly coupled - one should ignore any radiation
from this string (such as open string (glueball) radiation in the
bulk) that would be present in the AdS/CFT Correspondence since
that would be the wrong physics for QCD. The string then
encounters a hard wall in the geometry and spreads out along it
forming a traditional QCD-like string with energy growing with its
length. We have proposed that this string will break quickly once
there is sufficient energy in it to pair create a quark pair,
breaking the string. The two sub-strings then separate, radiating
energy through their end points as rho mesons. We expect the
majority of the hadronization particle production to come from
this radiation rather than repeated string breakings.

It is interesting to compare this picture to that of the Lund
string model \cite{Andersson:1983ia} which is one of the leading
descriptions of hadronization used in accelerator Monte Carlos.
The Lund model is based on the 1980s picture of the link between
QCD and string theory. The strings between quarks live in the 3+1d
of QCD. The model assumes that a very long string forms between
the quark anti-quark pair which then sequentially fragments with
the fragments being assigned as various hadronic states. It's
possible to morph our picture into the Lund one by assuming that
the D7 brane in Figure 2 lies very close to the hard wall - the
fifth dimension then plays little role and the string almost lies
in the same plane as the end points move. How long the string
grows and how many times it then breaks are not things we have
computed so we can not dispute the Lund model. The striking
difference though between the models is that significant rho meson
production occurs at the end points of the string moving in AdS,
removing the need for repeated breakings of the initial string. Of
course the differences in the AdS picture may be an artefact of a
large $N$ expansion and not relevant to true QCD, but equally
$N=3$ is believed to not be so far from $N \rightarrow \infty$.

In theories close to the ${\cal N}=4$ theory the rho mesons are
special, in that they are associated with operators whose
dimensions are protected from renormalization. This means they are
present as supergravity modes in the DBI action of the D7 brane -
other quark bound states would be represented by stringy states
also tied to the D7 world volume. In QCD we would not expect a
separation in character between these modes and all hadronic
species should be produced at the end point governed by the same
end point motion.

The idea of separating, radiating string fragments is reminiscent
of another reasonably successful model of hadronization. Thermal
models \cite{Becattini:1995if} have been proposed that treat the
event as separating fireballs radiating hadrons in thermal
equilibrium. Some discussion of how that thermal spectrum can
emerge from a gauge theory event with many final states even at
zero temperature can be found in
\cite{Hormuzdiar:2000vq,Hatta:2008qx} - it seems likely that the
core idea is just that the energy of hadronization is freely
available to all modes. A very toy model of holographic
hadronization was presented, based on these ideas, in
\cite{Evans:2007sf} - there an equal lump of energy was dumped
into the 5d holographic fields associated with each set of QCD
bound states with a Gaussian profile in the radial direction. The
model seemed to reproduce the data reasonably well too.
\footnote{Note to compare to data there is the need to impose an
elaborate decay chain between the initially produced particles and
the finally observed particles so the number of species for which
there is data is much less than the total number of species in the
initial yield which may hide many evils.} Here one can associate
that lump of energy to the wave of gauge field emitted by the
string end point as shown in Figures 9-11.

It's therefore interesting that our holographic model seems to
include aspects of both the Lund string model and thermal models
of hadronization. We hope that insights from AdS will lead to
phenomenologically more successful models of hadronization in QCD
in the future. \vspace{2cm}

\noindent {\bf Acknowledgements:} JF and ET would like to thank
STFC for their studentship funding. NE is funded by an STFC
rolling grant and is grateful to Marija Zamaklar and Kasper Peeters
for discussions. 
KJ would like to thank Andreas Karch for many
fruitful conversations. KJ was supported in part by the U.S.
Department of Energy under Grant No. DE-FG02-96ER40956.

\end{document}